# Plasmon–exciton coupling enhances second-order nonlinear response in borophene–ZnO hybrid structures

Maximilian Black[1,ǂ,*], Yaser Abdi[1,2,ǂ,*], Prabhdeep Singh[1,ǂ], Bharti Garg[1], Zahra Alavi[2], Mohammadreza Alikhanim[2], Mohammad Hossein Salemi Seresht[2], Fatemeh Chahshouri[1], Masoud Taleb[1,3], Nahid Talebi[1,3 *]

1. Institute of Experimental and Applied Physics, Kiel University, 24118 Kiel, Germany
2. Department of Physics, University of Tehran, 1439955961 Tehran, Iran
3. Kiel Nano, Surface, and Interface Science KiNSIS, Kiel University, 24118 Kiel, Germany
   * E-Mail: talebi@physik.uni-kiel.de, black@physik.uni-kiel.de, y.abdi@ut.ac.ir

ǂ These authors contributed equally to the paper.

**Abstract** – Nonlinear optical processes in low-dimensional materials are often weak or symmetry forbidden, limiting their use in nanoscale light sources and on-chip frequency conversion. Here, we show that combining two weakly nonlinear systems—anisotropic borophene and excitonic zinc oxide—yields an enhanced and resonant nonlinear response. In borophene–ZnO heterostructures, cathodoluminescence reveals a two-orders-of-magnitude enhancement at 400 nm and 800 nm, due to an enhanced two-photon absorption process. Under tunable near-infrared excitation, a clear second-harmonic signal emerges with quadratic power dependence and strong resonance near 800 nm. We attribute this to nonlinear plasmon–exciton coupling, which reshapes the excitonic response and enables efficient hybrid pathways for frequency conversion. These results establish anisotropic plasmon–exciton hybridization as a route to controlling nonlinear optical responses in low-dimensional heterostructures.

## Introduction

Nonlinear optics underpins a broad range of modern photonic technologies and spectroscopic tools, from ultrafast pump–probe and transient absorption spectroscopy to frequency comb generation and precision metrology[1-4]. Strong nonlinear optical responses are also central to laser-based nanofabrication[5], high-harmonic generation[6,7], and attosecond science[8,9]. However, efficient second- and higher-order nonlinear processes typically rely on bulky, intrinsically anisotropic crystals that support phase matching, imposing severe constraints on miniaturization and integration[10-12]. These limitations have motivated intense efforts to identify alternative routes toward enhanced nonlinear optical functionality in low-dimensional platforms.

A particularly promising direction exploits enhanced light–matter interactions at the nanoscale, where optical fields can be confined, reshaped, and amplified beyond what is possible in bulk media[13,14]. In this context, nonlinear exciton–plasmon interactions[15-17], phonon-mediated nonlinearities[18,19], and deeply non-centrosymmetric two-dimensional (2D) materials[20-22] have emerged as leading candidates for compact nonlinear photonic elements. Despite this progress, achieving strong and controllable second-order nonlinear responses in low-dimensional systems remains challenging.

Here, we demonstrate a hybrid nanoscale material system that overcomes these limitations by combining strong in-plane anisotropy with plasmon–exciton–mediated nonlinear light–matter interactions. Specifically, we show that interfacing $\chi_3$ borophene with excitonic ZnO nanorods elevates the nonlinear optical response of its individual constituents. The plasmon-exciton nonlinear interactions lead to a two-orders-of-magnitude enhancement in cathodoluminescence (CL) emission and enables highly efficient second-harmonic generation (SHG), with a measured second-order susceptibility of 2.3 $pmV^{-1}$ obtained under moderate peak laser intensities, generated from nanoscale geometries at single borophene sheets and ZnO nanorod interfaces.

Borophene is a synthetic two-dimensional material composed of a single atomic layer of boron[23-26]. Depending on its crystallographic phase, borophene exhibits a wide range of electronic and optical properties[27-29]. Among these, $\chi_3$ borophene is particularly notable for its pronounced in-plane anisotropy, displaying metallic and semiconducting behaviour along orthogonal crystallographic directions[30]. This anisotropy gives rise to an in-plane hyperbolic optical response, supporting highly directional polaritonic and plasmonic modes at wavelengths longer than 520 nm.

ZnO, by contrast, is a well-established direct-bandgap semiconductor with a bandgap of 3.37 eV at the Γ point and robust excitons with binding energies of approximately 60 meV, ensuring stability at room temperature. When synthesized in the wurtzite phase as nanorods, ZnO supports excitonic emission[31] along with pronounced emissions from defects[32], and exhibits moderate nonlinear optical responses under low peak excitation powers[33]. As we demonstrate here, hybridizing ZnO nanorods with $\chi_3$ borophene sheets fundamentally alters this behaviour. The CL emission from the hybrid system emerges at 400 nm and 800 nm and demonstrate a suppressed emission from defects. In addition, the second-harmonic response under resonant near-infrared light excitation is amplified compared to pure ZnO.

We attribute this pronounced nonlinear optical response to anisotropic plasmon–exciton coupling at the borophene–ZnO interface. Resonant interactions between ZnO excitons and interband transitions in borophene, combined with strong anisotropic plasmonic confinement and enhanced interfacial fields, give rise to efficient nonlinear polarization pathways. In particular, the highly anisotropic in-plane plasmonic response of borophene, together with its substantial optical absorption, increase light trapping efficiency and local field enhancement at resonance. These effects enable two-photon–mediated plasmon–exciton interactions that generate a strong second and third-order nonlinear polarization, thereby greatly increasing nonlinear CL-emission – originated from the two-photon absorption – and enhancing second-harmonic generation. These results establish anisotropic and nonlinear plasmon–exciton coupling as a powerful and global mechanism for activating nonlinear optical responses in low-dimensional materials at deep sub-wavelength scales.

## Results

The studied samples consist of $\chi_3$ borophene flakes synthesized via chemical vapor deposition (CVD), which were subsequently transferred onto ZnO nanorods using a wet-transfer method[34] (Figure 1a, Supporting Information). The ZnO nanorods were prepared through a hydrothermal reaction involving nitrate salts and hexamethylenetetramine, yielding high-aspect-ratio rods with well-defined wurtzite structure.

The $\chi_3$-phase borophene atomic structure is composed of a mixture of triangular and hexagonal motifs, with two in-plane principal axes oriented along the directions indicated by the x and y arrows (Figure 1b). This constitutes a highly anisotropic optical response, demonstrated by the permittivity components of the material along the principal axes (Figure 1d)[30]. The material exhibits a positive-valued permittivity component along the x-principal axis, where several transition peaks are apparent. Along the y-axis, the permittivity is Drude-like, showing a zero-crossing at 520 nm.

The absorption spectrum of borophene is highly polarization-dependent as well. By changing the polarization angle with respect to the x-principal axis and under the normal incident angle, the calculated absorption spectrum demonstrates a shift from semiconducting behaviour with four distinguished transitions at 420 nm, 517 nm, 643 nm, and 843 nm along the x direction to a broad and enhanced metallic absorption peak along y direction at longer wavelength (Figure 1c).

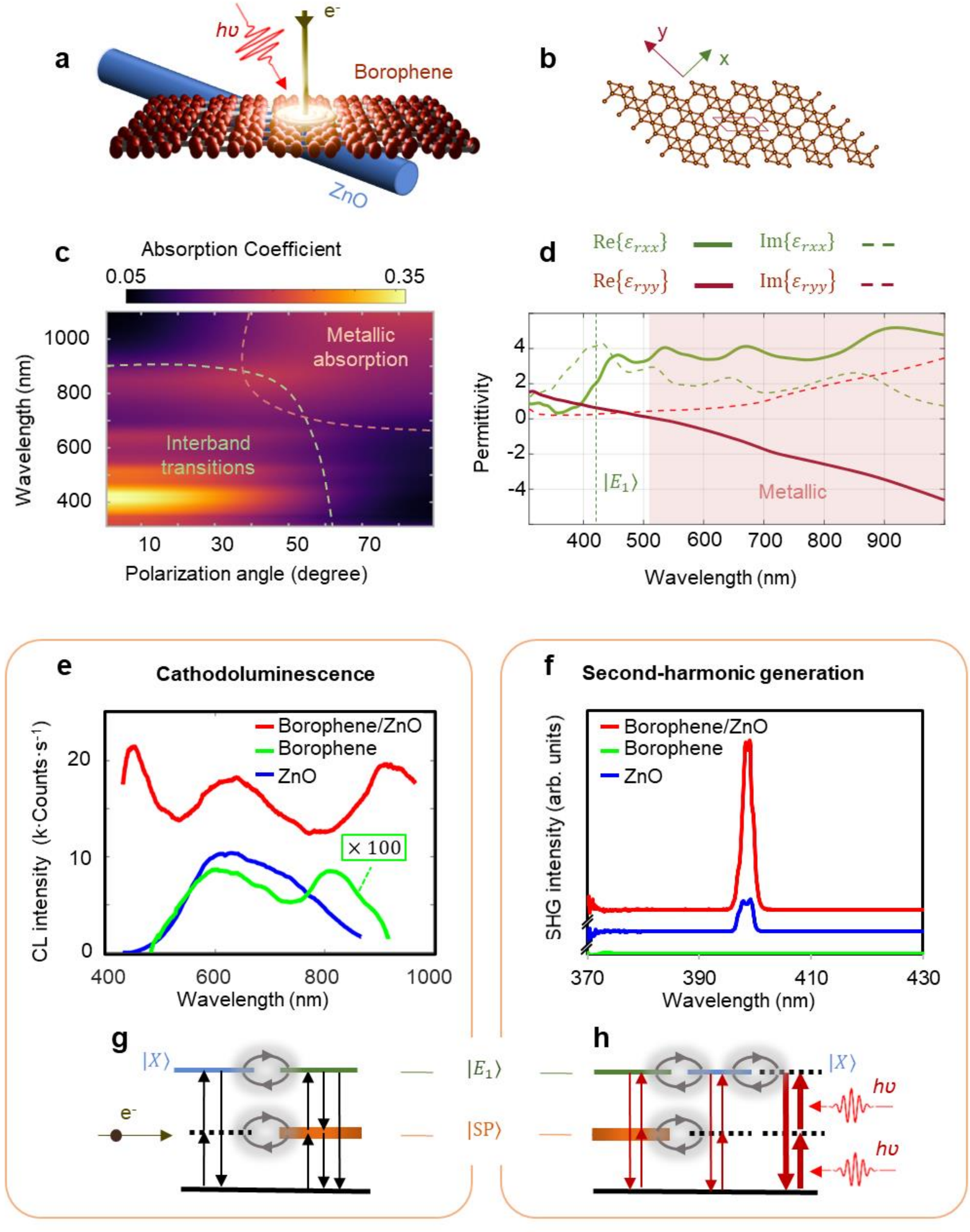

**Figure 1. Optical responses of hybrid borophene/ZnO structures.** **a**) Schematic of structure explored upon electron beam excitations and coherent laser excitation. **b**) The atomic structure of borophene, where x and y show the main crystallography axes. **c**) Calculated polarization dependent absorption spectrum of borophene under normal incident angle. **d**) Permittivity of borophene calculated along two principal axes. The solid and dashed lines show the real and imaginary components of permittivity along x-axis (green lines) and y-axis (red lines), respectively. **e**) CL response and **f**) SHG intensity of borophene (green line), ZnO nanorods (blue line), and Borophene/ZnO hybrid structures (red line). The SHG peaks becomes broadened in the measurement process while maximizing the collected signal. The excitation wavelength is 798 nm. Doubly-resonant excitation and emission pathways leading to **g**) enhanced CL response and **h**) enhanced SHG maps of the hybrid structure.

This interplay between semiconducting and metallic behaviour in borophene plays a significant role in the optical response of the ZnO/borophene hybrid structure (Figures 1e-h). We explore the optical response of this structure under electron-beam excitations as well as coherent and resonant laser excitations in the following. Using CL spectroscopy allows us to determine the nanoscale and local response of the hybrid structure with a high spatial resolution[35-39]. Furthermore, excitation by resonant and coherent tuneable supercontinuum laser excitation unravels the significant role of plasmon-exciton resonant and nonlinear interactions in generating second harmonic response.

***Cathodoluminescence spectroscopy***

First, we employed CL spectroscopy to investigate the optical properties of borophene, ZnO nanorods, and their hybrid heterostructure (Fig. 1e). Cathodoluminescence spectroscopy has established itself as a prominent technique for exploring incoherent and coherent material excitations in semiconductors and plasmonic systems, with high spatial resolution and large collection efficiency, when a parabolic mirror is used to collect the emitted light[36,38].

A comprehensive analysis of the morphological and structural properties of both the pure and hybrid systems was conducted, providing key insights into the crystallographic arrangement and interface formation (Supporting Information). The CL spectrum from pure borophene flakes reveals two broad and weak luminescence peaks, centered at 600 nm and 820 nm (Fig. 1e). The former is attributed to the luminescence caused by the broad continuum of interband transitions, where the latter is due to the surface plasmon excitation.

The CL response of ZnO when mediated by excitons is predicted to manifest a maximum at $\lambda = 386\ \mathrm{nm}$[40]. We observe the excitonic peak only in dense clusters of synthesized ZnO nanorods (Supporting Information), due to a higher density of excited excitons under electron beam irradiation that emits at a rate comparable to defects. The CL response of individual ZnO nanorods at the room temperature displays a broad emission centered at 650 nm (Fig. 1c), attributed to deep-level defects such as oxygen vacancies and interstitial zinc[41].

The CL response of the hybrid borophene/ZnO system (Fig. 1b) at the overlapping borophene and ZnO parts, however, reveals distinct sharper peaks centered at 850 nm and 420 nm (See Fig. 1c-e). The CL intensity is enhanced by two orders of magnitude compared to the CL emission from borophene. Compared to CL response of pure ZnO, the emission from defects does not uniquely determine the spectral features, but instead two enhanced and sharper peaks at 850 nm and 420 nm emerge. This emergence of a sharper and more pronounced peak at 850 nm and the new peak observed at 420 nm, i.e., approximately half of the main 850 nm peak, indicate incoherent two-photon absorption from the hybrid structure (Figure 1g). Nearly resonant excitonic response of the ZnO structure with the interband transitions in borophene at an energy level exactly two-times larger than the plasmonic resonances in borophene constitute multiple pathways for nonlinear absorption and remission in the hybrid borophene/ZnO structure.

The enhanced CL emission is repetitively observed in different hybrid borophene/ZnO structures, as well as the existence of suppressed emission from defects (Fig. 2a, b). In order to further explore the nonlinear anisotropic interaction between excitons and plasmons, we deposit a large high-quality borophene flake – as in contrast with the wrinkled flake in Fig. 2a – on top of multiple ZnO rods with varying orientations, as shown in the SEM image (Fig. 2c, d). A selected region of the hybrid structure was analyzed through CL hyperspectral imaging, uncovering distinctive and spatially localized emissions from only specific ZnO/borophene hybrids (Fig. 2f, g; compare the CL intensity from left and right rods marked by blue and red arrows respectively). The recorded spectra display emission at 400 nm, consistent with previous measurements – and a broad emission from defects. The CL emission is particularly pronounced at specific interface regions depending on the orientation of the rods, as revealed by hyperspectral imaging (Fig. 2e). These results are consistently repeated when large borophene sheets cover ZnO nanords with different orientations (Supporting Information).

Spectral analysis conducted along two nearly perpendicular orientations demonstrated that strong emission featured at 400 nm appears for rods oriented along one direction, whereas they were absent for the rods oriented along the orthogonal direction (Fig. 2f, g). For the latter rods, the CL emission is dominated not by plasmon-exciton interactions, but by the emission from defects in ZnO. To elucidate the origin of this anisotropic optical response, the crystallographic structure of borophene was mapped onto the hyperspectral image, revealing a correlation between nanorod orientation and borophene's intrinsic electronic anisotropy (Fig. 2d). Only when the ZnO rod is aligned with the metallic y-axis of borophene, the propagating plasmon polaritons along this direction can interact efficiently with ZnO excitons, accounting for a strong luminescence peak at longer wavelengths and the emergence of the two-photon absorption at approximately 400 nm. Moreover, this strong direction-dependent behavior exhibits the role of plasmons: Two plasmon-enhanced photon absorptions becoming resonant with excitonic resonances in ZnO and interband transitions in borophene lead to a direction-dependent and significantly enhanced CL emission. Moreover, the CL spectra of the borophene/ZnO hybrid structure was investigated for different electron beam currents and kinetic energies to elucidate the interfacial luminescence behavior (Supporting Information).

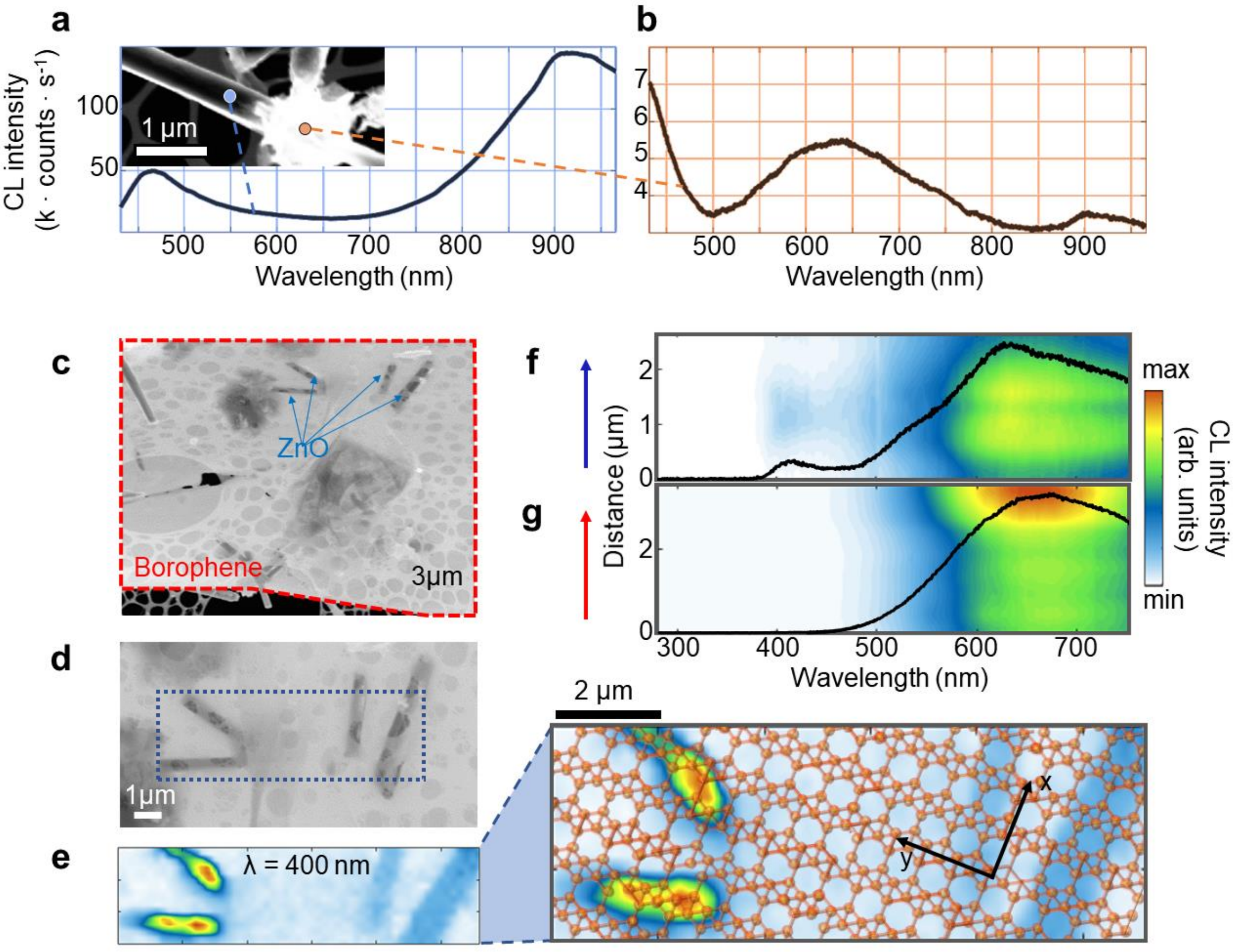

**Figure 2. CL response of borophene/ZnO structure. a,b)** CL spectra at depicted electron impact positions. The inset shows the SEM image of the structure composed of a wrinkled borophene sheet positioned on top of a ZnO nanorod. **c)** A large-view and **d)** a magnified SEM image of a large single-layer borophene sheet covering several ZnO nanorods. The dotted box demonstrates the area from which the hyperspectral CL images in (e) at the wavelength of 400 nm are represented. The right panel in **e)** exhibits an overlayed borophene structure with depicted crystalline configuration. CL spectra versus position along directions depicted by **f)** blue and **g)** red arrows on the SEM image in panel (d). Current and acceleration voltage were set to 5 nA and 10 keV, respectively.

***Optical spectroscopy***

Optical spectroscopy measurements further substantiate the nonlinear optical behavior of the hybrid system and the generation of a coherent second harmonic polarization in the hybrid borophene/ZnO nanorod systems (Fig. 3a). For exploring the second-harmonic generation under coherent laser excitation, a broad-range tunable laser system is used. A supercontinuum laser generates optical pulses with the temporal broadening of 100 ps, laser power of 2.3 mW, and the repetition rate of 5 MHz and it is spectrally filtered using an acousto-optical tunable filter to a bandwidth of below 2 nm. The laser is then focused onto the sample, separately using objectives with numerical apertures of NA=0.7 and Na=0.9. At wavelength of 798 nm, the borophene/ZnO hybrid structure generates emission precisely at 399 nm (Fig. 1f), which we identify as coherent SHG due its narrow band width. This effect is particularly pronounced at interfacial regions, where local field enhancement due to plasmonic confinement facilitates efficient frequency conversion (Fig. 3b to d). The luminescence mapping demonstrates that the SHG signal is spatially confined to interface regions, highlighting the critical role of heterointerfacial phenomena in enabling nonlinear optical responses (Fig. 3d). To verify whether the observed SHG enhancement originates from quasiparticle interactions at the interface and plasmon-exciton interactions, comparative measurements were performed on pure ZnO, pure borophene, and the Borophene/ZnO hybrid structure (Fig. 1f and Supporting Information).

In Fig. 3c, an SEM image displays a zinc oxide nanorod, partially coated with borophene. We focused a 798 nm laser pulse at specific points of the sample marked in this figure and measured its luminescence with a spectrometer. The optical images integrated over the wavelength range of our detector are dominated by the residual signal of the fundamental frequency of the laser (Fig. 3b). Light coupled out at the ends opposite to the focus position exhibits the excitation of a guided wave in the ZnO nanorod. Different spectra acquired by focusing the laser at depicted positions along the nanorod reveal that the emission is significantly stronger near the Borophene/ZnO interface (Fig. 3d). As the excitation laser moves away from the borophene-covered region into the ZnO nanorod, SHG intensity exhibits a pronounced decrease. Consistently, a detectable emission is observed even when the laser is focused at positions not covered by borophene, which is due to the non-centrosymmetric structure of pure ZnO nanorod yielding SHG alone.

To evaluate the SHG enhancement of the borophene/ZnO heterostructure, SHG emission was observed in pure ZnO rods with varying intensity depending on the position of the laser spot (Supporting Information). This is expected, since edges and corners of the rods break local crystallographic symmetry as well, resulting in an enhanced second-order nonlinear response. By comparing the SHG signal of a pristine ZnO rod with similar laser spot positions on ZnO rods in contact with borophene, we consistently obtain increased emission from the ZnO/borophene heterostructure. When illuminated on the borophene/ZnO interface, we observe a more than 2.7 times the intensity compared to the laser spot positioned on the same ZnO rod sufficiently far away from the borophene sheet (Supplementary Fig. 5b and c). Interestingly, the highest emissions were measured on Zno rods in the vicinity of borophene sheets. Here, we deduce that the heterostructure is still excited efficiently, while avoiding increased scattering through the wrinkled borophene. Further, this behavior is consistent with the CL measurements shown in Fig. 2a and b, where the electron impact positions near to borophene and not dierectly incident on borophene, enhances the two-photon absorption process.

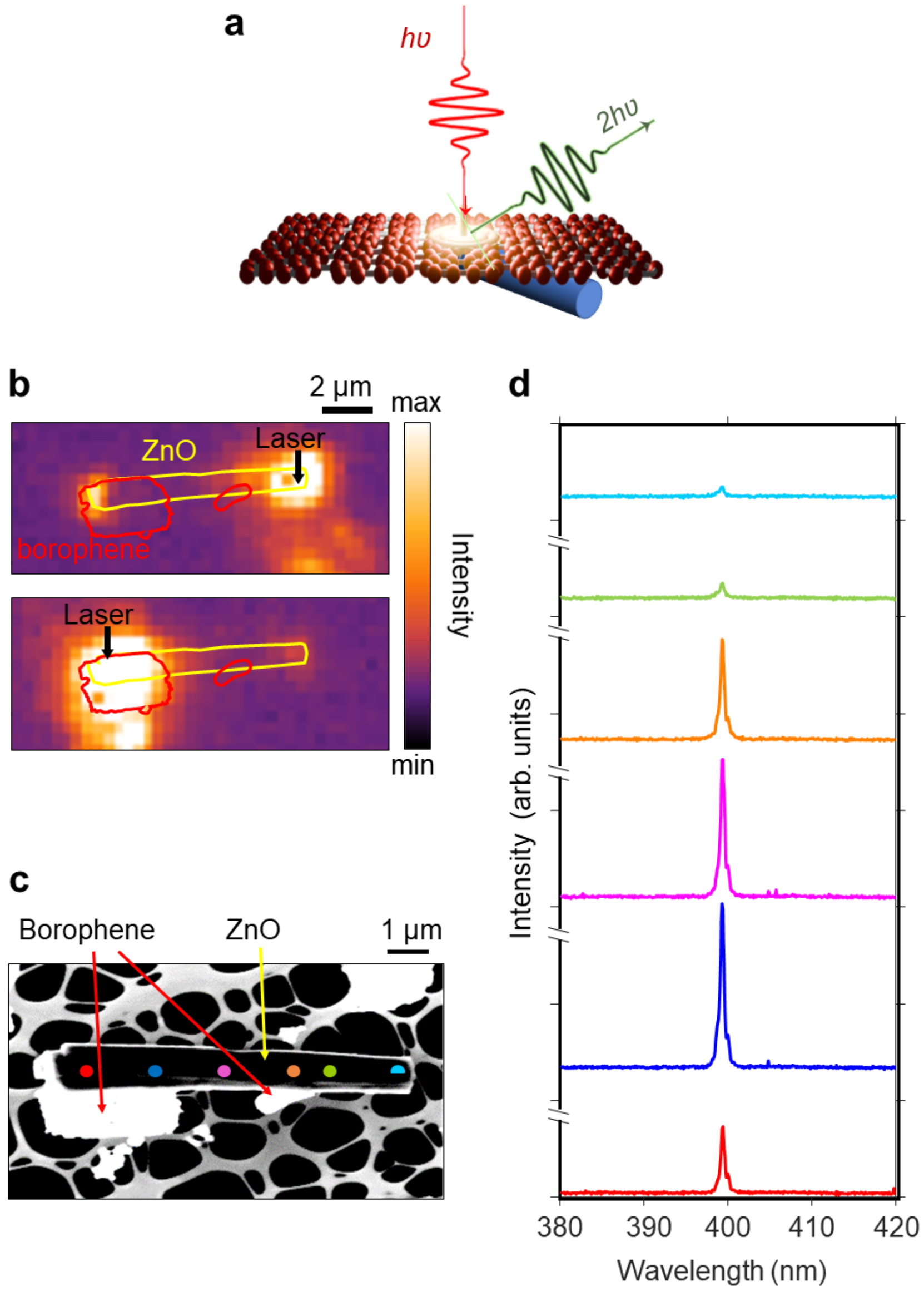

**Figure 3. Nonlinear optical response of Borophene/ZnO heterostructures trough optical spectroscopy. a)** Schematic of the experiment with the laser and emission of the second harmonic. **b)** Optical microscopy image integrated over wavelength. The fundamental frequency is filtered using a low pass filter, but its residual signal overlaps with the SHG signal, demonstrating enhanced scattering at the borophene positions. **c)** SEM image showing a ZnO nanorod partially coated with borophene. **d)** SHG spectra under 798 nm laser excitation from the positions marked in panel (c).

The optical measurements of Borophene/ZnO heterostructures reveal distinct nonlinear optical behaviors across different excitation wavelengths (Fig. 4a). The narrow-band emission spectra recorded under excitation wavelengths between 776 nm and 840 nm confirm the generation of strong nonlinear optical polarization at the Borophene/ZnO interface. Furthermore, the SHG peak wavelength follows the expected half-wavelength scaling of the excitation wavelength**,** consistent with second-harmonic generation process (Fig. 4b). The maximum SHG intensity occurs at 798 nm (Fig. 4a). In addition, in line with the second-order nature of the SHG process, the generated second harmonic power in the hybrid borophene/ZnO structure exhibits a quadratic dependence on the incident laser power (Fig. 4d). Note that the system investigated for the latter measurement consists of stacks of multiple ZnO-rods covered by borophene (Fig. 4c), potentially leading to further increased SHG signal, due an increased interaction volume.

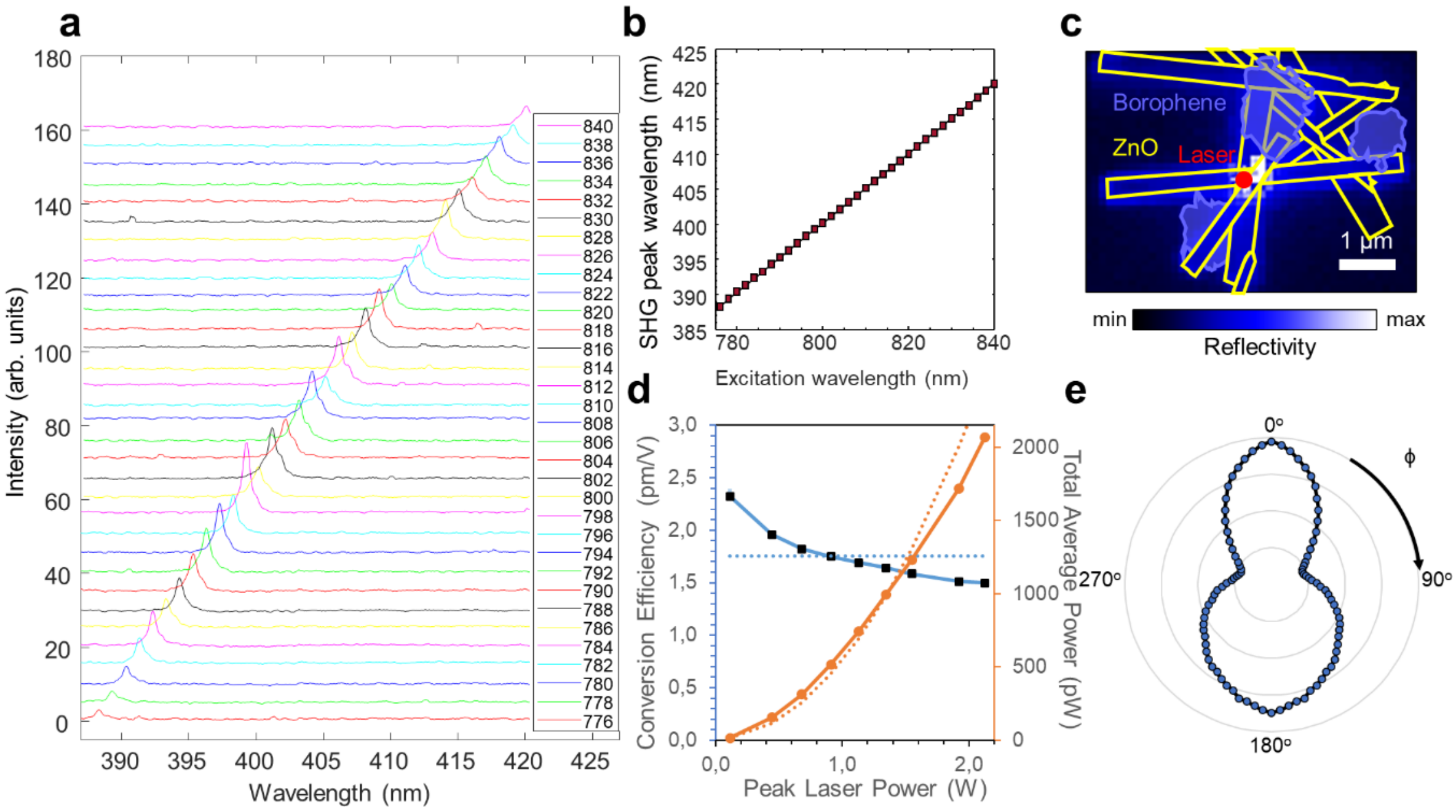

**Figure 4. SHG at the Borophene/ZnO heterointerface. a)** SHG spectra of borophene/ZnO under laser excitation in the wavelength range of 776–840 nm. **b)** SHG peak wavelength versus excitation wavelength following the expected half-wavelength scaling of the excitation laser. **c**) laser spot position on the borophene/ZnO hybrid system used for the power and polarization dependent SHG measurements. **d**) Quadratic dependence of SHG power on incident laser power, confirming second-order nonlinear behavior (orange line). **e**) Polarization-dependent SHG measurements reveal pronounced optical anisotropy.

The polarization-dependent SHG measurements of the Borophene/ZnO heterostructure reveal pronounced optical anisotropy (Fig. 4e). The SHG signal exhibits variations in peak intensity as the laser polarization changes, confirming that the nonlinear optical response is strongly dependent on the polarization orientation, since the SHG efficiency depends on the orientation of the incident electrical field with respect to the c-axis of the nanorod.

The electric field of the second harmonic signal exhibits a quadratic dependence on the incident electric field. Hence, second-order susceptibility, which we use as a measure for the SHG conversion efficiency, is defined as

$$\chi_2 = \sqrt{\frac{2}{Z_0}}\sqrt{\frac{P_{2\omega}}{A_{2\omega}}} \cdot \frac{A_\omega}{P_\omega}, \quad (1)$$

where $P_{2\omega}$ and $P_\omega$ are the emitted powers at the second-harmonic and fundamental frequencies, $A_{2\omega}$ and $A_\omega$ are the respective emission and illumination areas, and $Z_0$ is the vacuum impedance.

By assuming $A_{2\omega} = A_\omega$, we obtain an average SHG conversion efficiency of $\chi_2 = 1.754\ \mathrm{pm/V}$. We observe a decrease of $\chi_2$ with rising laser power, however, this is an effect of a subsequent departure from the optimized settings for position and focus during the measurement. Hence, the best value is obtained at a laser peak power of 0.113 W with $\chi_2 = 2.32\ \mathrm{pm/V}$.

The SHG conversion efficiency measured from pure ZnO nanorods in our system shows a pronounced dependence on the excitation position, laser spot size, and the geometrical configuration of the nanorods (Supporting Information). The maximum value observed in our measurements, obtained when exciting near the nanorod edge, corresponds to an effective nonlinear susceptibility of $\chi_2 = 1\ \mathrm{pmV^{-1}}$. Reported SHG efficiencies for ZnO nanostructures vary widely in the literature, ranging from $\chi_2 = 0.1\ \mathrm{pmV^{-1}}$ to $15\ \mathrm{pmV^{-1}}$, depending on nanorod dimensions, excitation and detection geometries, and substrate[40,41].

The efficiency of the incoherent electron-induced two-photon absorption process is strikingly stronger than the coherent SHG process, since the density of excitons generated by electron beams is significantly larger than that generated by optical excitation. Upon electron irradiation, a large number of secondary or backscattered electrons are produced in the ZnO rods, which sequentially excite many excitons[42]. Therefore, these sequential interactions give rise to an incoherent two-photon absorption process rather than coherent SHG.

**Conclusion**

In summary, we have observed two-photon absorption and SHG in borophene/ZnO hybrid nanostructures, explored using cathodoluminescence and optical spectroscopy techniques. The plasmon-exciton interactions at the heterointerface, combined with borophene's anisotropic optical properties and ZnO's wide-bandgap excitonic response, lead to an exceptionally efficient nonlinear optical process, observable even in the cathodoluminescence signal. Moreover, the spatial confinement of SHG at interfacial regions and its polarization-dependent characteristics underscore the critical role of nonlinear excitonic-plasmonic coupling in shaping nonlinear light-matter interactions at the nanoscale.

Beyond its fundamental significance, these findings pave the way for leveraging low-dimensional van der Waals heterostructures in next-generation nonlinear photonics. The ability to engineer enhanced nonlinear responses through interfacial quasiparticle interactions in tailored heterostructures provides a promising route toward ultrafast optical signal processing and on-chip nonlinear photonic functionalities. By enabling strong nonlinearities in compact, low-dimensional platforms, such mechanisms may further support frequency conversion schemes relevant for emerging quantum photonic technologies. By exploiting cathodoluminescence-based nonlinear characterization, this study introduces an advanced tool for probing plasmon-exciton hybridization, offering unique spatial resolution and insight into nanoscale nonlinear optical phenomena. These results establish borophene/ZnO as a compelling platform for nonlinear optics and inspire future explorations into tailored heterostructures for tunable nonlinear photonic applications, particularly further enhanced via integrating them in photonic cavities and waveguides.

**Methods and characterization techniques**

*Sample Preparation*

Borophene sheets were synthesized via chemical vapor deposition (CVD) and subsequently transferred onto desired substrates and TEM grids using a wet transfer method, as detailed in our previous studies[30]. This approach ensures high-quality exfoliation while preserving structural integrity.

ZnO nanorods were hydrothermally synthesized by mixing $NO_3^{2-}$ and hexamethylenetetramine (HMTA) at 7 mM in a 1:1 ratio. The solution was incubated at 90°C for 24 hours in a foil-covered container to promote controlled nucleation and growth. After synthesis, the ZnO nanorods were mechanically detached from the substrate and suspended in DI water for enhanced dispersion. They were then carefully transferred onto holey carbon grids, ensuring uniform spatial distribution.

To construct the heterostructure, borophene sheets were subsequently integrated onto the ZnO nanorods via a wet transfer approach. This process enabled precise control over borophene coverage while maintaining the distinct properties of both components. The final heterostructure was prepared for characterization with optimized sample integrity.

*Cathodoluminescence Spectroscopy*

Cathodoluminescence (CL) measurements were performed in a field-emission scanning electron microscope (Zeiss Sigma) equipped with a DELMIC SPARC detection system. The optical emission generated by the electron–matter interaction was collected using an aluminum-coated parabolic mirror mounted above the sample inside the SEM chamber. The sample was positioned at the focal point of the mirror to maximize the photon collection efficiency. The mirror collects the emitted radiation over a large solid angle and redirects it toward the optical detection path for spectral analysis, enabling spatially resolved CL spectroscopy of the sample.

During CL measurements, the focused electron beam acts as a broadband excitation source, inducing optical emission through electronic and optical excitations in the material. To minimize electron-beam-induced damage, the acquisition parameters—including beam current, spot size, scanning step size, and exposure time—were systematically optimized prior to data collection. For spectral measurements, the electron beam scanning step sizes were varied between 5 and 20 nm depending on the measurement. An exposure time of 200 ms per pixel was used to maintain a balance between signal intensity and surface integrity.

To acquire the CL spectra over the broad spectral range shown in Figures 1e, 2a, and 2b, a long-pass filter with a cutoff wavelength of 700 nm was used to suppress artifacts arising from second-order diffraction of the spectrograph gratings around 800 nm. The reported spectra are composite (patched) spectra, obtained by combining measurements recorded without the filter in the range of 460–700 nm and measurements acquired with the long-pass filter in the range of 700–960 nm.

The acquired CL datasets were processed using the ODEMIS software environment for spectral preprocessing and noise reduction. The processed spectra were subsequently exported in CSV format and further analyzed in MATLAB for data visualization and quantitative analysis.

*Optical Spectroscopy*

To measure the SHG signal, a supercontinuum laser source (SC-OEM YSL) was utilized and spectrally filtered using an acousto-optical tunable filter (AOTF-PRO-D). This setup enabled precise wavelength selection for the excitation of borophene/ZnO hybrid samples. The samples were illuminated through a Nikon Eclipse Ti2-A inverted microscope, which also collected the emitted photons. For the site-selective SHG measurement in Figure 3 and for the wavelength-dependent measurement NA = 0.7 objective was used to focus the laser, while the rest of the measurements were performed with a NA = 0.9 objective.

To effectively isolate the SHG emission from the excitation light reflected along the optical path, dichroic mirrors were employed in conjunction with edgepass filters. This optical arrangement ensured spectral separation and minimized background interference.

Spatially resolved optical spectra were recorded using a Princeton Instruments SpectraPro HRS 500-S grating spectrometer equipped with a reflective grating of 600 grooves/mm, providing high spectral resolution for accurate detection and analysis. The slit of the spectrometer was opened to collect the whole emitted signal. This leads to an increased SHG intensity in exchange for spectral resolution, resulting in a broadening of the SHG peaks.

*Structural and Morphological Characterization*

Lattice structure and selected area electron diffraction (SAED) patterns of borophene sheets were investigated using a Philips CM300 transmission electron microscope operated at 120 kV. High-resolution imaging provided insights into crystalline ordering and defect structures.

Morphological characterization of the heterostructures was performed using a Zeiss-SIGMA field-emission scanning electron microscope (SEM), enabling topographical and size distribution analyses.

To determine the crystal structure, X-ray diffraction (XRD) measurements were conducted using a Philips-PW1730 instrument with a wavelength of 1.54 Å, an operational current of 30 mA, and a voltage of 40 kV. This technique provided quantitative phase identification of the samples.

**Acknowledgement**

This project has received funding from the European Research Council (ERC) under the European Union’s Horizon 2020 research and innovation programme (grant agreements no. 101170341, Kiel, UltraSpecT), the Deutsche Forschungsgemeinschaft (DFG) under grant numbers 525347396 and 447330010, and by the VolkswagenStiftung through a Momentum Grant.

**Data availability:** The data are available from the corresponding authors on reasonable request.

# Supplementary Information

## Plasmon–exciton coupling enhances second-order nonlinear response in borophene–ZnO hybrid structures

Maximilian Black[1*], Yaser Abdi[1,2], Prabhdeep Singh[1], Bharti Garg[1], Zahra Alavi[2], Mohammadreza Alikhani[2], Mohammad Hossein Salemi Seresht[2], Fatemeh Chahshouri[1], Masoud Taleb[1,3], Nahid Talebi[1,3 *]

1. Institute of Experimental and Applied Physics, Kiel University, 24118 Kiel, Germany
2. Department of Physics, University of Tehran, 1439955961 Tehran, Iran
3. Kiel Nano, Surface, and Interface Science KiNSIS, Kiel University, 24118 Kiel, Germany

* E-Mail: talebi@physik.uni-kiel.de, black@physik.uni-kiel.de, y.abdi@ut.ac.ir

**Content:**

## 1. Structural and morphological analysis of the borophene/ZnO

The structural and morphological analysis of the borophene/ZnO hybrid system confirms the successful integration of borophene sheets onto ZnO nanorods. Scanning electron microscopy (SEM) imaging reveals the deposition of borophene layer on ZnO nanorods, forming a hybrid architecture with significant interfacial interaction (Supplementary Fig. 1a). The sample preparation follows a two-step process, wherein borophene is synthesized via chemical vapor deposition (CVD) and subsequently coated onto dispersed ZnO nanorods on a substrate.

Further insights into the structure are provided by bright-field transmission electron microscopy (TEM), which illustrates the physical coupling between borophene and ZnO nanorods, highlighting their distinct contrast and layered features (Supplementary Fig. 1b). Complementary X-ray diffraction (XRD) analysis verifies the crystallographic nature of the ZnO component, exhibiting characteristic diffraction peaks corresponding to the wurtzite phase (Supplementary Fig. 1c). Due to the ultra-thin nature and limited coverage of borophene, its diffraction signature remains undetectable in the XRD spectra.

To investigate the crystallinity at the nanoscale, selected-area electron diffraction (SAED) patterns of both borophene and ZnO provide direct evidence of their respective lattice structures. The SAED profile of ZnO presents well-defined diffraction spots consistent with the wurtzite phase (Supplementary Fig. 1e), while borophene exhibits distinct diffraction features attributed to its characteristic atomic arrangement (Supplementary Fig. 1d). These findings collectively substantiate the structural coherence of the hybrid system, laying the foundation for its distinctive optical and electronic properties.

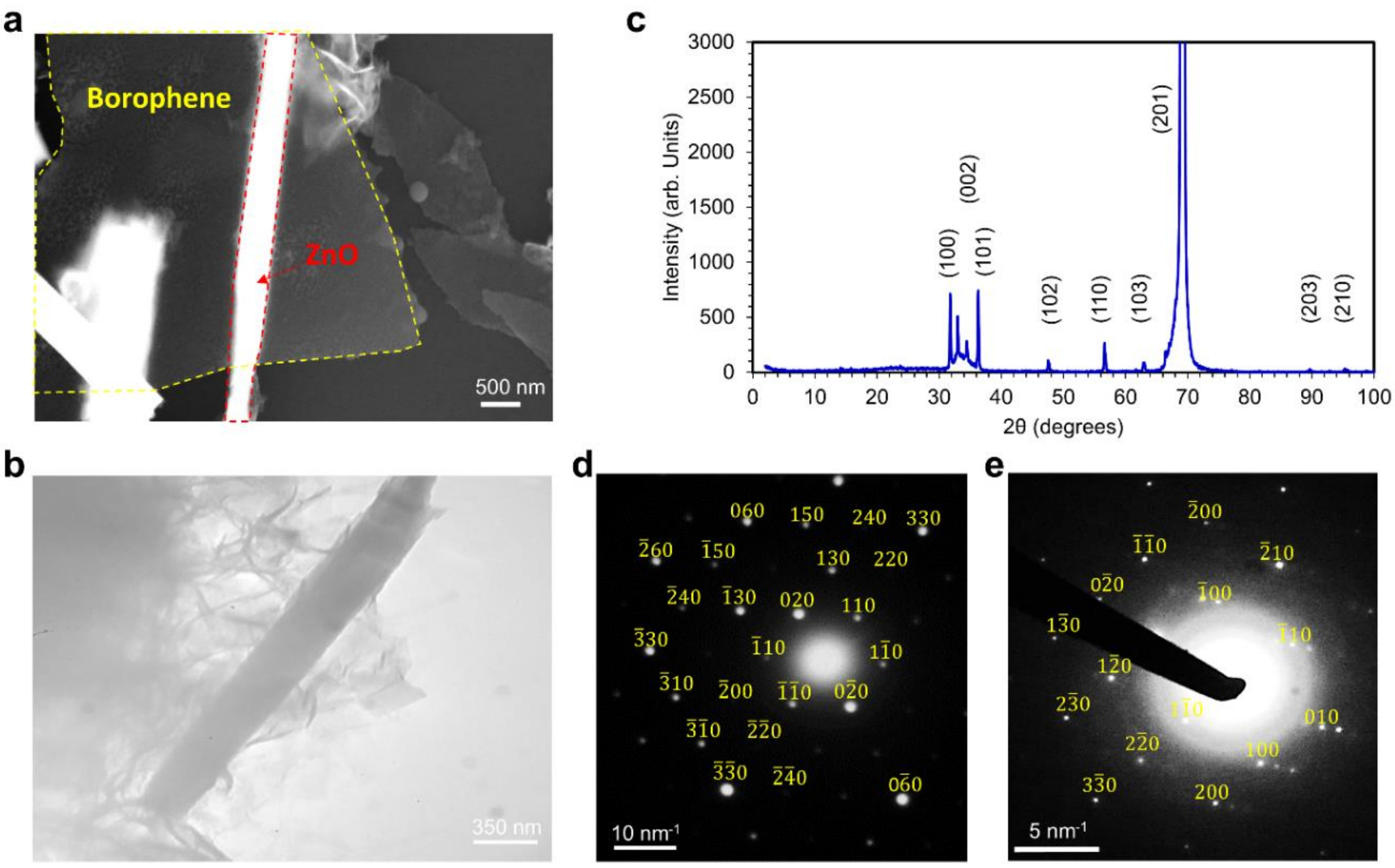

**Supplementary Fig. 1 | Structural and morphological characterization of borophene/ZnO hybrid structures. a**, SEM image showing borophene sheets deposited on ZnO nanorods. **b**, Bright-field transmission electron microscopy revealing the architecture of the hybrid system. **c**, X-ray diffraction spectrum of ZnO, with indexed diffraction peaks confirming the wurtzite phase; no significant diffraction from borophene is observed due to its low coverage. **d**, SAED pattern of borophene, highlighting characteristic diffraction spots corresponding to its atomic lattice. **e**, SAED of ZnO, illustrating distinct crystallographic features of the wurtzite structure.

## 2. Optical absorption spectra of borophene

The polarization dependent absorption spectrum of borophene is calculated as below. We assume a borophene layer positioned at $z = 0$ plane, with its crystalline axes oriented along x and y axis as shown in Fig. 1b of the main text. The borophene sheet is excited with a plane wave at the normal incident angle with respect to the z=0 plane. The electric field distributions at domains specified by $z > 0$ and $z < 0$ are constructed as

$$\vec{E}(z>0) = \left(E_0 \cos\varphi \hat{x} + E_0 \sin\varphi \hat{y}\right) e^{-ik_0 z} + \left(R_x E_0 \cos\varphi \hat{x} + R_y E_0 \sin\varphi \hat{y}\right) e^{+ik_0 z} \quad (1)$$

and

$$\vec{E}^2(z<0) = \left(T_x E_0 \cos\varphi \hat{x} + T_y E_0 \sin\varphi \hat{y}\right) e^{-ik_0 z}, \quad (2)$$

respectively. Here $\varphi$ is the polarization angle of the incident electric field, $E_0$ is the electric field amplitude, $k_0$ is the free-space wavenumber of the light, $R_x$, $R_y$, $T_x$, and $T_y$ are the polarization-dependent electric-field reflection and transmission coefficients. The magnetic field is obtained using $\vec{H} = (i\omega\mu_0)^{-1} \vec{\nabla} \times \vec{E}$. The borophene is treated as conductive layer with a negligible thickness. Therefore, the boundary conditions are provided as $E_x^1(z=0) = E_x^2(z=0)$, and $\hat{z} \times \left(\vec{H}_1 - \vec{H}_2\right) = \sigma_{xx} E_x \hat{x} + \sigma_{yy} E_y \hat{y}$, with $\sigma_{\alpha\alpha} = -i\omega\varepsilon_0\left(\varepsilon_{r,\alpha\alpha} - 1\right)$, and $\alpha = x, y$. Equations (1) and (2) together with the boundary conditions lead to $T_\alpha = \left(1 + \frac{1}{2}\eta_0\sigma_{\alpha\alpha}\right)^{-1}$ and $R_\alpha = T_\alpha - 1$. The power transmission and reflection coefficients are calculated using the Poynting theorem as $r_p = |R_x|^2 \cos^2\varphi + |R_y|^2 \sin^2\varphi$ and $t_p = |T_x|^2 \cos^2\varphi + |T_y|^2 \sin^2\varphi$, and the absorption coefficient is obtained as $a_p = 1 - r_p - t_p$.

## 3. Cathodoluminescence response of a dense ZnO nanorod cluster

The cathodoluminescence spectra of pristine ZnO nanorods are dominated by defect-related emission, reflecting the high density of intrinsic and surface states typical for nanoscale ZnO. Only in dense assemblies of nanorods and nanostructures does a comparatively weak near-band-edge excitonic emission become discernible. This behavior is in stark contrast to that observed in borophene/ZnO heterostructures, where the CL response is governed by excitonic emission, while defect-related channels are strongly suppressed.

Moreover, in the hybrid system the excitonic resonance of ZnO is markedly shifted as a result of strong plasmon–exciton interactions at the borophene/ZnO interface. This nonlinear coupling renormalizes the exciton energy and brings the emission wavelength to around 400 nm, where it becomes resonant with the interband transitions of borophene, thereby enabling highly efficient radiative recombination.

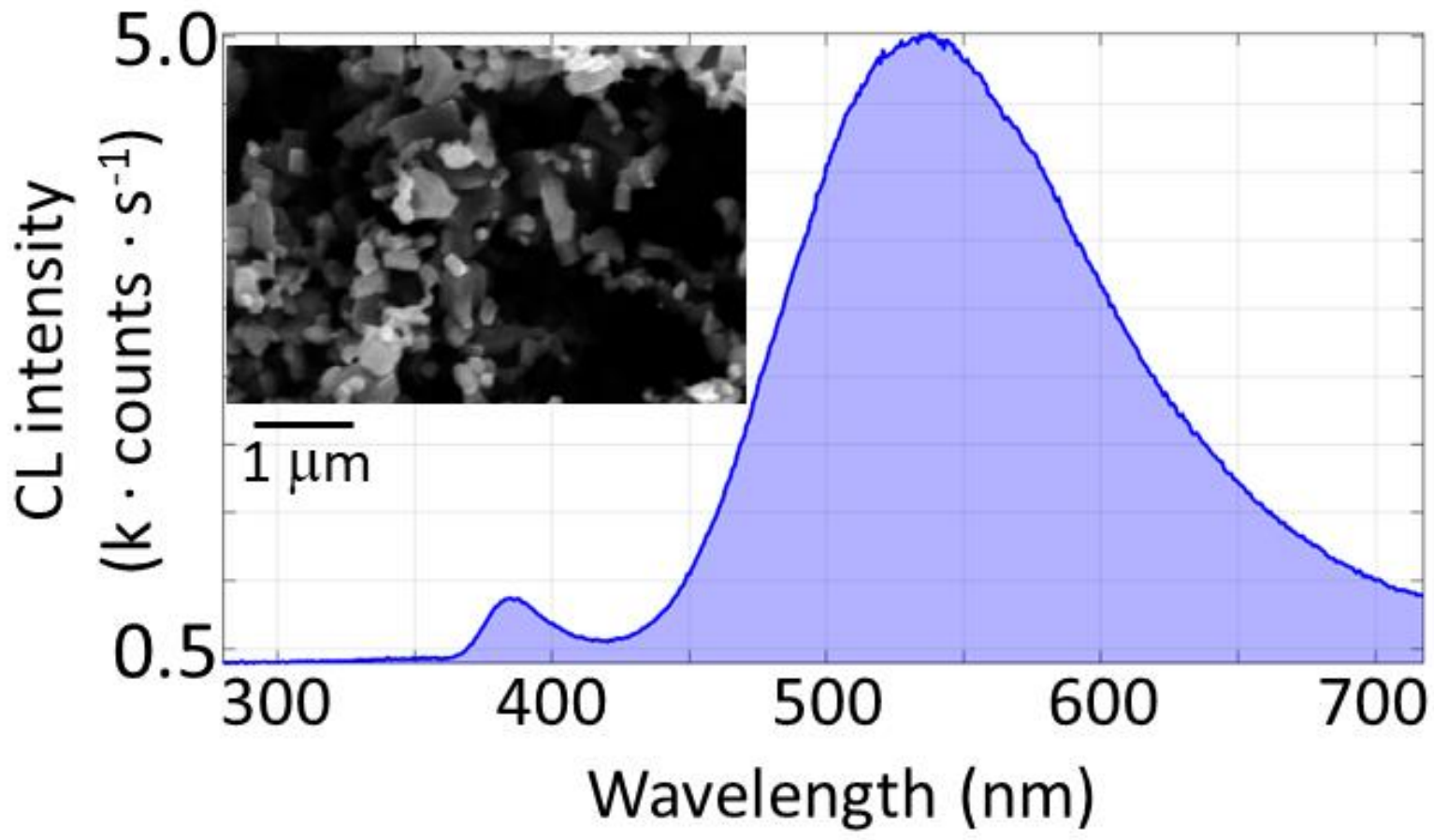

**Supplementary Fig. 2 | Cathodoluminescence response of a dense cluster of ZnO nanorods.** The CL spectrum exhibits an excitonic peak at $\lambda = 386\ \mathrm{nm}$ and a broad spectral resonance at $\lambda = 540\ \mathrm{nm}$ due to the emission from defects. The inset exhibits the SEM image of the ZnO nanorod cluster.

## 4. Cathodoluminescence response of borophene/ZnO hybrid structure

Figure 2 of the main text demonstrates that the orientation of the ZnO nanorod covered by borophene strongly affects the observed cathodoluminescence (CL) signal. In particular, the intensity of the spectral peak associated with the two-photon absorption process exhibits a pronounced dependence on the nanorod orientation, reflecting the anisotropic optical response of borophene. Specifically, borophene displays metallic behavior primarily along its *y* crystallographic axis (see Fig. 1b and d). This anisotropy enables efficient coupling between plasmons in borophene and excitons in ZnO when the symmetry axis of the ZnO nanorod is aligned with the *y*-axis of the borophene layer.

Supplementary Fig. 3 provides an additional example illustrating this anisotropic coupling between differently oriented nanorods (SEM image shown in Supplementary Fig. 3a) and the borophene layer. Both the hyperspectral CL images (Supplementary Fig. 3b and c) and the spectra acquired across the nanorods (Supplementary Fig. 3d) reveal that the CL intensity at λ = 420 nm is negligible for Rod 2 (marked by a blue arrow), whereas it reaches a maximum for Rod 1 (marked by a red arrow). Notably, these two nanorods are oriented approximately perpendicular to each other, aligning with the *x* and *y* crystallographic axes of the borophene layer, respectively.

## 5. Cathodoluminescence of borophene/ZnO under varying electron beam conditions

The interaction between exciton and plasmon quasiparticles is strongly localized to the interface between the ZnO nanorods and the borophene flake. Using electron beams with different kinetic energies, the excitation can be tailored to either excite the interface or the bulk of the nanorods (Supplementary Fig. 4). In addition to the kinetic energy, the current of the electron beam influences the acquired signal significantly.

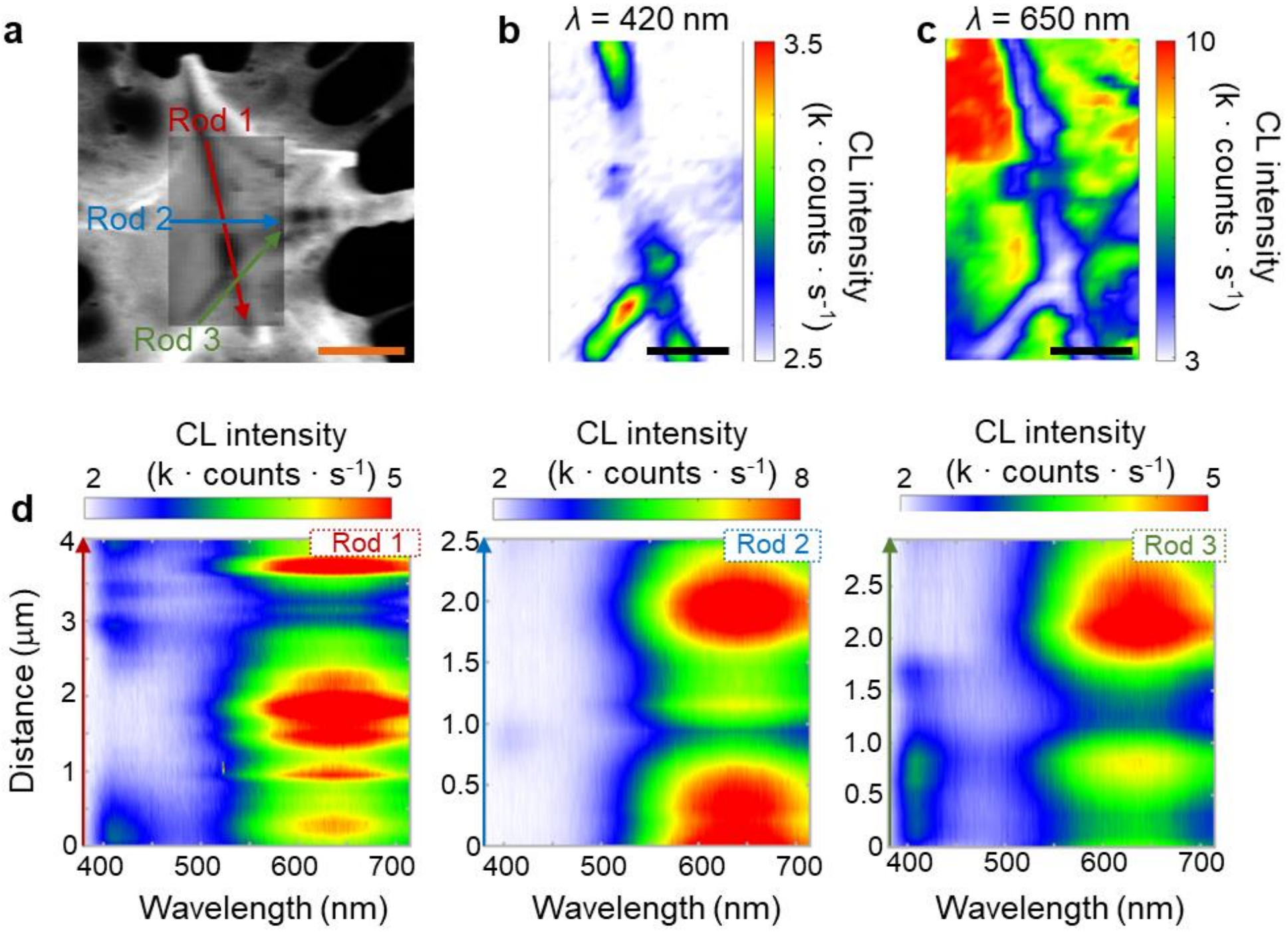

**Supplementary Fig. 3 | Cathodoluminescence response of a large borophene flake covering multiple ZnO rods with different orientations. a)** SEM image of the explored structure. Scale bar is 2 µm. **b**,**c)** hyperspectral images at depicted wavelenths. Scale bar is 1 µm. d) CL spectra versus position along the directions highlighted by arrows on the SEM image.

We particularly observe that lower acceleration voltages allow to better configuring the nonlinear peak associated to the two-photon absorption process. The largest signal associated with this peak is observed for the Kinetic energies below 10 keV and beam currents above 7 nA. Noticeably, we do not observe any radiation damage at voltages as high as 20 keV and currents as large as 10 nA, demonstrating the structural stability of borophene even under extreme excitation conditions.

Two distinct regions were analyzed: One where borophene is deposited onto ZnO nanorods and another where the electron beam excites the nanorod at positions not covered by borophene. The dependence of CL intensity on EHT voltage further reveals the critical role of the heterointerface in luminescence efficiency. Higher voltages lead to deeper electron penetration into the ZnO, reducing excitation efficiency at the borophene/ZnO interface. Consequently, at the acceleration voltages of 15 kV and 20 kV, the emission spectra closely resemble those of bare ZnO. In contrast, lower voltages of 5 and 10 kV yield enhanced peak intensities. These findings strongly indicate that the borophene/ZnO interface is the primary contributor to the observed CL features, reinforcing the significance of interfacial quasiparticle interactions in defining optical responses.

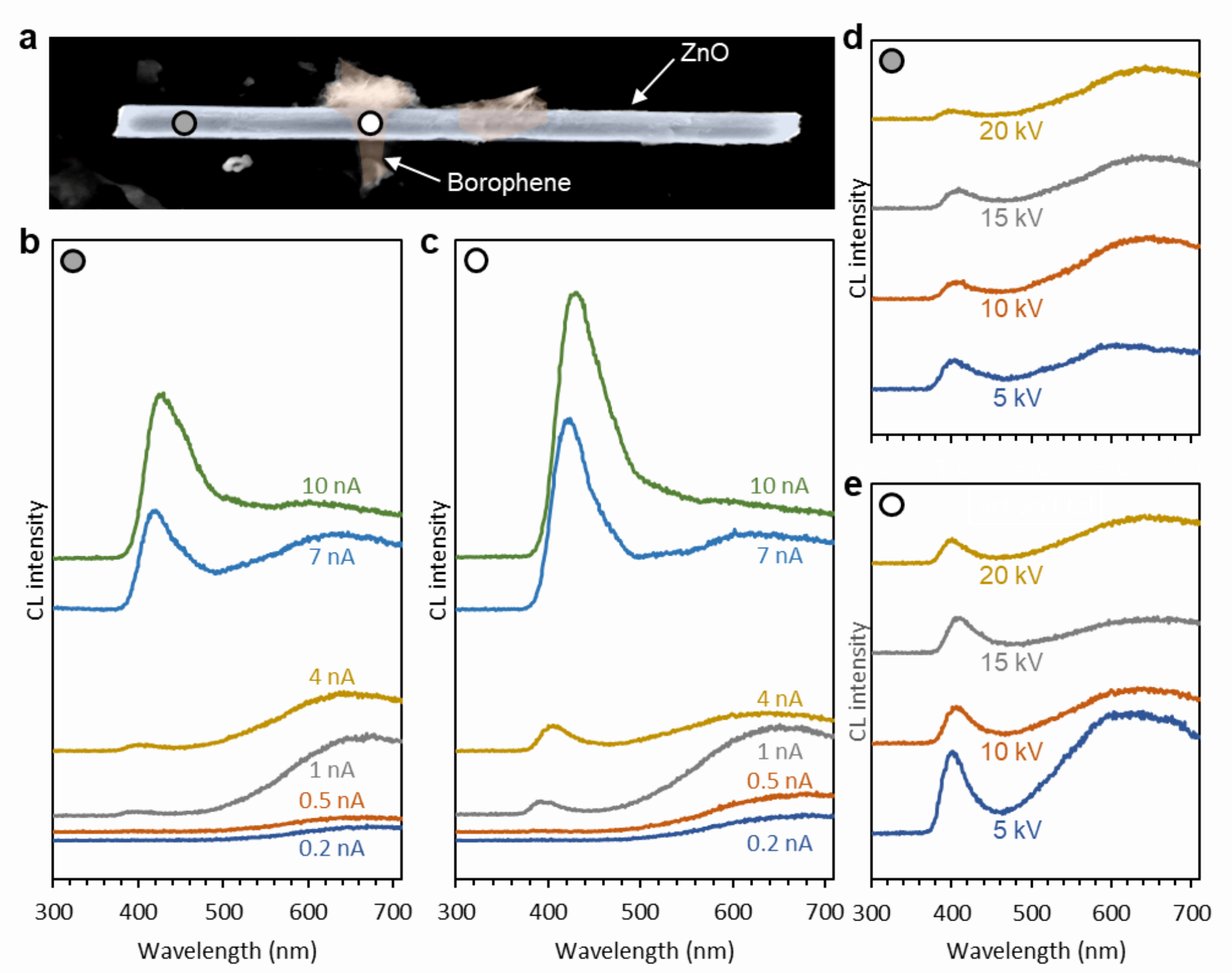

**Supplementary Fig. 4 | Electron beam parameter-dependent cathodoluminescence in borophene/ZnO hybrid structures. a)** SEM image of the borophene/ZnO hybrid structure, highlighting two distinct regions: the interfacial area (red circle), where borophene overlays ZnO nanorods, and the pristine ZnO region (blue circle). **b,c)** Cathodoluminescence spectra recorded at varying electron beam currents, ranging from 200 pA to 10 nA. The interfacial region exhibits intensified emission features with this enhancement becoming more pronounced at higher currents (notably at 4 nA). Additionally, the spectra reveal a redshift in luminescence with increasing beam current. **d,e)** Cathodoluminescence spectra obtained under different electron beam voltages at the current of 7 nA, demonstrating minimal impact on luminescence intensity in the pristine ZnO region, while the interfacial region shows considerable variations in spectral intensity. At higher voltages (15, 20 kV), electron penetration into the ZnO reduces excitation efficiency at the interface, resulting in spectra resembling those of bare ZnO. In contrast, at lower voltages (5, 10 kV), interfacial excitation is enhanced, confirming that the heterointerface is the dominant contributor to these spectral features.

## 6. Dependence of the Second Harmonic signal on the excitation location

The efficiency of the generated second-harmonic (SHG) signal strongly depends on the excitation position, even within a single nanorod or heterostructure. Supplementary Fig. 5 shows representative examples in which the laser beam is focused at different locations along the nanorod, exciting either the borophene-covered ZnO nanorod, regions adjacent to the borophene flake, or parts of the nanorod far from the heterostructure interface. We consistently observe a higher SHG conversion efficiency when the laser is focused near the borophene flake, but not directly on it. This behavior can be attributed to coupling to the optical modes of the ZnO nanorod, where the excitation position determines the coupling efficiency and becomes optimal when the laser overlaps with local field hot spots of these modes. In contrast, direct excitation of borophene leads to stronger absorption due to its metallic nature, which reduces the effective coupling to the optical modes of the ZnO nanorod and consequently decreases the SHG efficiency.

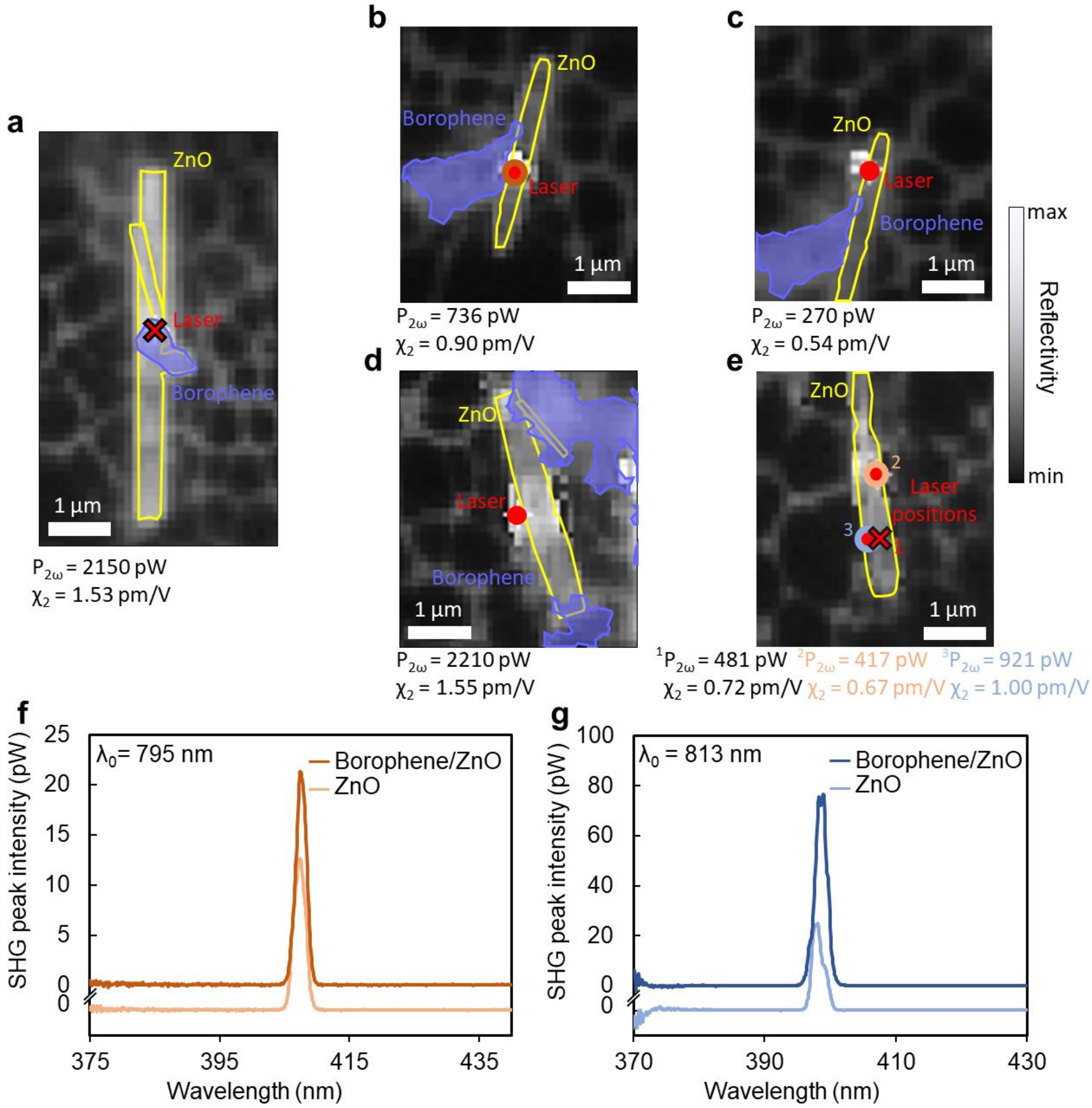

**Supplementary Fig. 5 |** Comparison of SHG signals acquired at various positions in the Borophene/ZnO heterostructure and in pure ZnO rods. **a to e**) Reflection images of ZnO rods and Borophene/Zno hybrid systems under combined brightfield and laser illumination together with the measured peak SHG intensities and second-order susceptibilities. a) Laser spot position marked by a cross on the borophene/ZnO hyprid system used for Figure 1f. **b**) SHG signal acquired when exciting the system directly on the borophene/ZnO interface as compared to the SHG in **c**), with the laser spot on the ZnO rod being far away from the borophene. **d**) High SHG at the edge of a ZnO rod close to the borophene sheet. **e**) Various laser positions on a pristine ZnO rod. **f**) SHG spectrum of the measurements on Borophene/ZnO marked by a cross in (a) compared to the position on pure ZnO marked by a light-orange coloured dot. **g**) Weakest measured SHG on borophene/ZnO marked by a dot in (d) compared to the strongest measured SHG marked by a light-blue dot in (e).

We note that the microscope objective limits the angular range of the collected radiation. To state a proper estimate for the fraction of the collected power to the total power, we consider an in-plane dipole radiation pattern and integrate over the solid angle that corresponds to the numerical aperture of the objective, resulting in 32.62%. While considering this would increase our SHG signal and $\chi_2$ by factors of 3.06 and 1.75, respectively, the exact radiation pattern remains unknown, hence these factors are not taken into account in the values presented in this work. Importantly, our detection configuration is not optimized for SHG measurements at 400 nm, as the spectral response of our CCD detector decreases significantly near this wavelength, leading to a conservative estimate of the absolute efficiency. For this reason, we benchmark the SHG response of the borophene/ZnO heterostructures directly against that of pristine ZnO nanorods measured under identical experimental conditions. Within this controlled comparison, the hybrid structures exhibit a consistently enhanced SHG response, which we attribute to the double-resonant plasmon–exciton coupling at the heterointerface.